\documentstyle[pmart,twoside,fleqn,epsf]{article} 
\newcommand{\be}{\begin{equation}}
\newcommand{\ee}{\end{equation}}

\begin{document}            
\title{MARSHALL-PEIERLS SIGN RULE IN FRUSTRATED HEISENBERG CHAINS}

\author{Andreas Voigt and Johannes Richter}

\address{Institut f\"ur Theoretische Physik, 
Otto-von-Guericke-Universit\"at Magdeburg\\ Postfach 4120, 39106
Magdeburg, Germany}

\date{July 30, 1999}

\maketitle                   
\pacs{75.10.Jm,75.40.Mg,75.50.Ee}

\begin{abstract}
We consider the frustrated antiferromagnetic s=1 Heisenberg quantum
spin chain with regard to the Marshall-Peierls sign rule (MPSR).  By
using exact diagonalization data we investigate the breakdown of the
MPSR in dependence on frustration and compare our findings with data
for s=1/2. We calculate a critical value of frustration $J_2^{crit}$
where the MPSR is violated. The extrapolation of this value to the
infinite chain limit holds $J_2^{crit} \approx 0.016$, lower than in
the case of s=1/2 ($J_2^{crit} \approx 0.027$). This points to a
stronger influence of frustration in the case of s=1.  Nevertheless the
calculation of the weight of the Ising-states violating the MPSR shows
that the MPSR holds approximately even for quite large frustration and
may be used for numerical techniques.
\end{abstract}

\section{Introduction}
\label{intro}

The Marshall-Peierls sign rule (MPSR) determines the sign of the
Ising-basis-states building the ground-state wave function of a
Heisenberg Hamiltonian~\cite{marshall55} and has been proven exactly
for bipartite lattices and arbitrary site spins by Lieb, Schultz and
Mattis~\cite{lieb61}. As pointed out in several papers the knowledge of
the sign is of great importance in different numerical methods, e.g.
for the construction of variational wave functions
\cite{retzlaff93zpb}, in Quantum Monte-Carlo methods (which suffer from
the sign problem in frustrated systems
\cite{readt81}) and also in the density matrix renormalization group
method, where the application of the MPSR has substantially improved
the method in a frustrated spin system \cite{schollwoeck98}.

The MPSR has been analyzed so far for systems with s=1/2. The authors
of  \cite{zeng95} studied the frustrated chain and found
for the ground-state a critical value for the breakdown of the MPSR for
the infinite chain limit by using exact diagonalization data. For the
$J_1$-$J_2$ model on the square lattice the violation of the MPSR was
considered as an indication for the breakdown of long range order
\cite{richter94epla,bishop98ccm}. In a recent paper \cite{voigt97pa} we
extended these investigations to higher subspaces of $S^z$. For linear
chains we have shown that for the lowest eigenstates in every subspace
of $S^z$ there is a finite region of frustration where the MPSR holds.

In this paper we want to analyze the frustrated spin chain with s=1.
This spin system has attracted a lot of attention, because of the well
known Haldane conjecture \cite{haldane88}. The unfrustrated s=1 spin
chain shows a spin gap and exponential decaying correlations whereas
the s=1/2 spin chain has no gap and a power-law correlation decay.
Since both systems are qualitatively different one might expect also a
different influence of frustration on the MPSR.

\section{The Model and the Marshall-Peierls sign rule}
\label{model_rule}

In the following we study the MPSR for the frustrated antiferromagnetic
s=1 Heisenberg quantum spin chain:
 
\be
\label{H12}
{\rm \hat H} =J_1 \sum_{\langle {\bf NN} \rangle} {\bf s_i} {\bf s_j} 
+ J_2 \sum_{\langle {\bf NNN} \rangle} {\bf s_i} {\bf s_j},
\ee

$\langle {\bf NN}\rangle$ and $\langle {\bf NNN}\rangle$ denote
nearest-neighbor and next-nearest-neighbor bonds on the linear chain.
We set $J_1=1$ for the rest of the paper. For this model the MPSR can
be exactly proved only for $J_2 \le 0$.

The Marshall-Peierls sign rule can be described as follows: In the
unfrustrated limit of $J_2=0$, the lowest eigenstate of the Hamiltonian
(\ref{H12}) in each subspace of fixed eigenvalue $M$ of the spin
operator $S_{total}^z$ reads
\be
\label{GZ:Phi}
\Psi_{M} = \sum_{m}{c_{m}^{(M)}|m \rangle} \hspace{0.2cm},
\hspace{0.5cm} c_{m}^{(M)}>0 \hspace{0.2cm}.
\ee
Here the Ising-states $|m\rangle$ are defined by
\be
\label{mpsr}
|m\rangle \equiv (-1)^{S_A-M_A}|m_1\rangle \otimes |m_2 \rangle \otimes
\cdots \otimes |m_N\rangle \hspace{0.2cm},
\ee
where $|m_i\rangle,\hspace{0.2cm}i=1,\cdots,N$, are the eigenstates of
the site spin operator $S_{i}^{z}$ ($ -s_{i} \leq m_{i} \leq s_i$),
$S_A= \sum_{i\in A}s_i$, $M_{A(B)}=\sum_{i\in A(B)}m_i$, $M=M_A+M_B$.
The lattice consists of two equivalent sublattices $A$ and $B$.
$s_i\equiv s$, $i=1,\cdots,N$, are the site spins. The summations in
Eq.(\ref{GZ:Phi}) are restricted by the condition $\sum_{i=1}^N m_i=M$.
The wave function (\ref{GZ:Phi}) is not only an eigenstate of the
unfrustrated Hamiltonian ($J_2=0$) and $S_{total}^z$ but also
of the square of the total spin ${\bf S}_{total}^2$ with quantum number
$S=\mid \!M \! \mid$. Because $c_m^{(M)}>0$ is valid for each $m$ from the
basis set (\ref{mpsr}) it is impossible to build up other orthonormal
states without using negative amplitudes $c_m^{(M)}$. Hence the
ground-state wave function $\Psi_M$ is nondegenerated. As it comes out,
the MPSR is still fulfilled not only for the ground-state but also for
every lowest eigenstate in the subspace $M$ in the unfrustrated case.
We emphasize that for $J_2>0$ no proof for the above statements can be
given and that a frustrating $J_2>0$ can destroy the MPSR.

\section{Results}
\label{resul}

\begin{figure}[t]                           
  \epsfysize=6cm           
  \centerline{\epsfbox{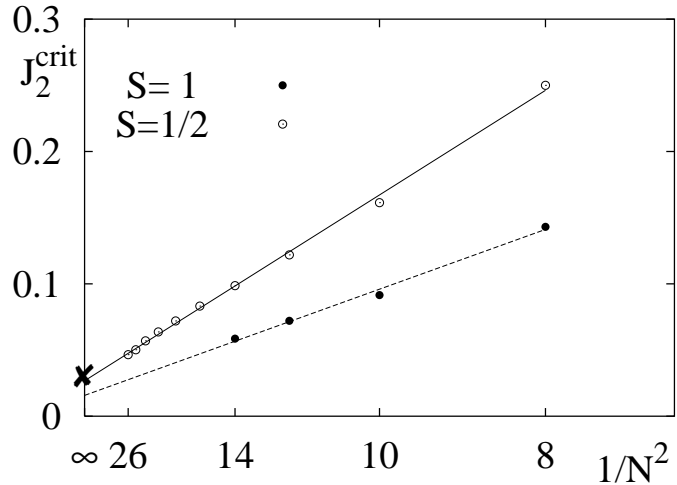}}    
  \caption{The critical value of frustration $J_2^{crit}$, where the
           MPSR starts to be violated as a function of the system size
           N. The cross denotes the value of Zeng and Parkinson
           \protect{\cite{zeng95}}.} 
  \label{fig1}                     
\end{figure}

We have calculated by exact diagonalization the ground-state of the
model (\ref{H12}) for N=8,$\ldots$,14 varying the frustration parameter
$J_2$. By analyzing the ground-state wave function according to the
MPSR we found for every system a critical value of frustration
$J_2^{crit}$, where the MPSR starts to be violated. We apply the
scaling law proposed by Zeng and Parkinson \cite{zeng95} and
extrapolate our data as a function of 1/N$^2$. We found a value for the
infinite chain limit: $J_2^{crit}(\infty)=0.016\pm0.003$. In
Fig.\ref{fig1} we compare these data with the values for the s=1/2
systems (N=8,$\ldots$,26), where the extrapolation yields
$J_2^{crit}(\infty)=0.027\pm0.003$. It is also interesting to note that
this value is slightly lower than the value of 0.032 found by Zeng and
Parkinson \cite{zeng95} using data for N=8,$\ldots$,20 only. 

\begin{figure}[t]                           
  \epsfysize=6cm           
  \centerline{\epsfbox{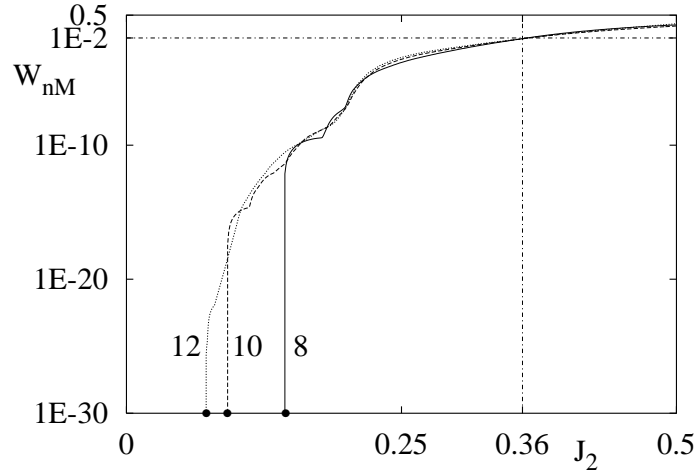}}    
  \caption{The weight of non-Marshall-states of the ground-state
           wave function $W_{nM}$ as a function of
           frustration $J_2$ for systems with N=8,10,12.}
  \label{fig2}                     
\end{figure}

We argue that in the case of s=1 the chain is more sensitive to
frustration and therefore the MPSR is violated for smaller values of
$J_2$. Nevertheless in numerical methods the MPSR can be used at least
approximately for much larger values of frustration. This can be
justified by the examination the ground-state wave function according
to the Ising-basis-states which violates the MPSR. We call these states
non-Marshall-states and denote their weight by $W_{nM}$. In
Fig.\ref{fig2} we show $W_{nM}$ as a function of frustration $J_2$.

As can be seen in Fig.\ref{fig2}, the weight of the non-Marshall-states
remains smaller than 1\% (1E-2) until $J_2 \approx 0.36$. This result
seems to be more or less size independent because all three lines for
the systems with N=8,10 and 12 cross at this point. The points at the
bottom line denote the first violation of the MPSR in a given system
and coincide with the points given in Fig.\ref{fig1}. The examination
of $W_{nM}$ indicates that for quite large frustration the predominant
part of the ground-state wave function fulfills the MPSR. Therefore,
the MPSR can be used in numerical methods even if it does not hold
strictly.

\section{Conclusions}
\label{concl}

We have shown that in the frustrated antiferromagnetic s=1 Heisenberg
quantum spin chain the Marshall-Peierls sign rule is violated by
frustration. We found by extrapolation to the infinite chain limit a
critical value of frustration $J_2^{crit} \approx 0.016\pm0.003$ below
which the MPSR still holds exactly. By calculating the weight of the
Ising-basis-states of the ground-state wave function which do not
fulfill the MPSR we conclude that the MPSR can be used in numerical
methods at least approximately until a large frustration of $J_2
\approx 0.36$.

\section*{Acknowledgments}
We would like to thank Nedko Ivanov for many useful discussions. This
work has been supported by the DFG (Project Nr. Ri 615/6-1).

\end{document}